\documentclass[11pt]{article}

\usepackage[final]{acl}
\usepackage{times}
\usepackage{latexsym}
\usepackage[T1]{fontenc}
\usepackage[utf8]{inputenc}
\usepackage{microtype}
\usepackage{inconsolata}
\usepackage{booktabs}
\usepackage{amsmath}
\usepackage{array}
\usepackage{graphicx}
\usepackage{ragged2e}

\usepackage{xurl}

\usepackage{siunitx} 

\title{Auditing Source Exposure in Baidu and Google AI Search}

\author{
Yibo Li \quad Enci Guan \quad Yuedan Cai \quad Geng Liu \quad Francesco Pierri\\
Politecnico di Milano, Italy \\
\texttt{\{yibo2.li, enci.guan, yuedan.cai\}@mail.polimi.it}\\
\texttt{\{geng.liu, francesco.pierri\}@polimi.it}
}

\begin{document}
\maketitle

\begin{abstract}
AI-generated overviews are becoming an increasingly prominent layer of search interfaces, yet their behavior in Chinese-language search remains underexplored. We conduct a cross-lingual audit of AI overview behavior on Baidu and Google using English queries sampled from MS MARCO and their translated Chinese counterparts. Our analysis examines when overviews are triggered across platform-language settings, which host domains receive visible exposure in Chinese-language overviews, how concentrated that exposure is, and how source overlap varies across settings. We also compare the embedding-based semantic similarity of generated answers for matched query intents. The results reveal substantial differences across platform-language settings in overview availability and visible source exposure. At the aggregate level, the settings exhibit low overlap in visible host-domain inventories, while matched-query answers yield median cosine similarities ranging from \num{0.701} to \num{0.813}. These findings indicate that answer-level semantic similarity and aggregate source exposure capture distinct dimensions of AI-mediated search. Evaluations of AI search should therefore consider not only the content of generated answers but also how source visibility is distributed across platforms, languages, and information environments.
\end{abstract}

\section{Introduction}

Search engines have moved from document-ranking interfaces toward answer-producing interfaces~\cite{kirsten-etal-2026-characterizing,grossman2026generativeaidisruptssearch}. Instead of returning only a ranked list of webpages, search result pages may now present an AI-generated overview above or alongside traditional search results.  Such overviews typically summarize an answer and display a small set of visible source references~\cite{liu2023evaluatingverifiabilitygenerativesearch}. As a result, they function not only as summary components but also as a new layer of information selection~\cite{huang2026answerbubblesinformationexposure}. This shift raises an important measurement question. In traditional search, source visibility is largely determined by ranking position and  the resulting user interaction~\cite{10.1145/1076034.1076063}. AI overviews redistribute this visibility by displaying a selected set of sources alongside generated answers. Auditing AI search therefore requires examining not only the generated answer but also when overviews appear, which sources they reference, and how concentrated this source exposure is ~\cite{aral2026riseaisearchimplications,grossman2026generativeaidisruptssearch}. This matters because AI overviews mediate not only what information users read, but also which sources remain visible in the search interface. Differences in source exposure may therefore shape which websites and information providers users encounter, even when generated answers are semantically similar.


Existing empirical audits of generative search have focused primarily on Google and other English-dominant systems. Using user-side observations, these studies examine overview triggering, source
selection, and citation exposure
\citep{kirsten-etal-2026-characterizing,
grossman2026generativeaidisruptssearch,
hu2026auditinggooglesaioverviews,
aral2026riseaisearchimplications}. However, comparatively little is known about these patterns in Chinese-language search or how they vary across platforms and languages.


This gap is particularly relevant in cross-lingual settings, where query language may be associated with different web infrastructures and platform ecosystems. Queries expressing the same information need in different languages may therefore lead search systems to expose different sources, including domestic platforms, encyclopedic sources, institutional websites, social media, and commercial websites. Prior work has shown that source exposure in AI-mediated search can vary across systems, languages, and query formulations~\cite{huang2026answerbubblesinformationexposure,chen2025generativeengineoptimizationdominate}.

We examine AI overviews as observable search-interface outputs, focusing on Chinese-language search. Rather than inferring the internal retrieval or generation mechanisms of Baidu and Google, we analyze whether an overview appears, how many source references it displays, which host domains receive visibility, and how source exposure varies across platform-language settings. Our primary comparison is between Baidu Chinese and 
Google Chinese, with Google English providing a cross-lingual reference.
Baidu English is included only in the overall analysis of overview appearance because it produced too few successful cases for reliable downstream source analysis. Consequently, the study does not provide a fully crossed design
for estimating independent platform and query-language effects.
We therefore interpret the results as differences among the
observed platform--language settings rather than as isolated
effects of platform or language.

We address three research questions:

\begin{itemize}
    \item \textbf{RQ1:} How frequently do AI overviews appear on Baidu and Google under Chinese- and English-query settings?

    \item \textbf{RQ2:} Which host domains are displayed in Chinese-language AI overviews on Baidu and Google, and how concentrated is their exposure?
    
    \item \textbf{RQ3:} How do visible host-domain overlap and answer-level
semantic similarity vary across Baidu Chinese, Google Chinese,
and Google English?
\end{itemize}

To address these questions, we collect \num{1998}\footnote{
The intended dataset contained \num{2000} observations, corresponding to \num{500} queries across four platform-language settings. One observation was not recorded in each Google setting, resulting in \num{499} collected records for Google Chinese and \num{499} for Google English. These missing observations were therefore not assigned to the Success, NoOverview, Blocked, or Error categories.} query-platform
observations from Baidu and Google. The base query set is drawn
from MS MARCO~\cite{bajaj2018msmarcohumangenerated}, a large-scale search-oriented dataset constructed from real, anonymized Bing queries.
MS MARCO provides realistic user information needs and a common set of search intents for aligned comparisons across platforms and languages. We retain the original English
queries and translate them into Simplified Chinese, producing two approximately aligned language versions of each query intent. We analyze overview appearance across all four platform-language settings. 
Topic-level and source-oriented analyses focus on Baidu Chinese and Google Chinese, with Google English serving as a cross-lingual reference. We measure overview appearance, source volume, host-domain concentration, cross-setting host-domain overlap, and query-level semantic similarity between generated answers. 

Overall, our findings show that different search settings can display  semantically similar answers while displaying  substantially different sets of visible host domains. 

\section{Related Work}

\paragraph{Generative search and source exposure.}
Generative search systems synthesize retrieved information into
answer-like outputs, creating a distinct layer of source selection
and visibility. Studies of Google Search, Gemini, AI Overviews, and
Featured Snippets show that generative search differs from traditional
retrieval in its synthesis, stability, and presentation of sources
\citep{kirsten-etal-2026-characterizing,
grossman2026generativeaidisruptssearch,
hu2026auditinggooglesaioverviews}.
AI-mediated search can also reshape which sources are exposed to users
and affect downstream website traffic
\citep{aral2026riseaisearchimplications,
huang2026answerbubblesinformationexposure,
khosravi2026impactaisearchsummaries}.
Much of this literature focuses on Google and other Western or
English-dominant systems, leaving Chinese-language search comparatively
underexplored.

\paragraph{Citation support and source concentration.}
Prior work examines whether citations in generative search adequately
support generated claims and how source exposure is distributed across
websites \citep{liu2023evaluatingverifiabilitygenerativesearch,
yang2025newssourcecitingpatterns}.
Other studies caution that visible citations do not necessarily
guarantee reliable, high-quality, or non-synthetic evidence
\citep{allaham2026syntheticsourcesauditinggenerative,
10.1145/3715275.3732089}.
Our study does not assess claim-level support or source quality;
instead, it treats visible host-domain frequencies as an
interface-level measure of source exposure and concentration.
\begin{figure*}[t]
    \centering
    \includegraphics[width=\textwidth]{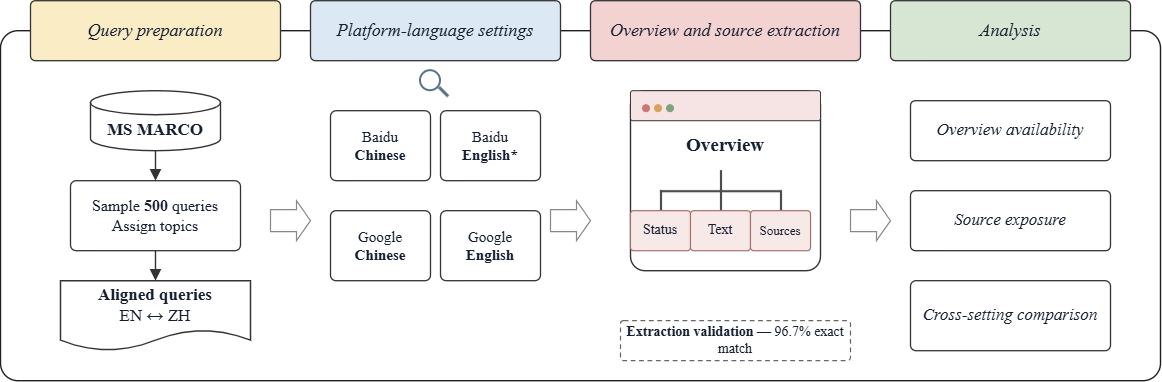}
    \caption{Overview of the audit pipeline. English queries are selected from MS MARCO and automatically translated into Simplified Chinese. The aligned queries are submitted to Baidu and Google, after which AI overviews, answer texts, and visible source domains are extracted for downstream analysis. Baidu English is included only in the overall overview appearance analysis.}
    \label{fig:audit-pipeline}
\end{figure*}
\paragraph{Cross-platform and cross-lingual search.}
Comparisons of Baidu and Google have identified platform- and
language-dependent differences in autocompletion, moderation, and
factual answers
\citep{liu2024comparisonautocomplete,
liu2026reliabilityasymmetries}.
Related work on generative engine optimization further examines how
content and source characteristics affect visibility in generated
search responses
\citep{chen2025generativeengineoptimizationdominate,
10.1145/3637528.3671900}.
Building on these studies, we conduct a user-side audit of overview
triggering, host-domain concentration, source overlap, and answer-level
similarity across Baidu Chinese, Google Chinese, and Google English.

\section{Data and Methods}

Figure~\ref{fig:audit-pipeline} summarizes the data collection and analysis workflow, from query selection and translation to overview collection, source extraction, and metric computation.

\subsection{Query Selection and Translation}

The query set is drawn from MS MARCO v1.1, a search-oriented dataset containing real anonymized user queries originally submitted to Bing~\citep{bajaj2018msmarcohumangenerated}. We choose MS MARCO because its queries reflect realistic information needs and provide a common set of query intents that can be compared consistently across platforms and languages. After removing duplicates, we randomly sampled \num{500} queries without topic-based filtering to preserve the natural distribution of user information needs. We retain the original English queries and translate them into  Simplified Chinese using Google Translate through the \texttt{deep\_translator}\footnote{
\url{https://pypi.org/project/deep-translator/}
} Python package. This produces aligned English and Chinese versions of each query. To assess translation quality, one author, a native Chinese
speaker proficient in English, manually reviewed a random sample
of 100 English--Chinese query pairs. A translation was judged
acceptable when it preserved the primary information need of the
original English query without substantially adding, omitting, or
altering the requested information. All 100 sampled translations
were judged acceptable. Before data collection, each English query was assigned exactly one of six topic labels using ChatGPT-5.5\footnote{
\url{https://help.openai.com/en/articles/11909943-gpt-5-5-in-chatgpt}
} and a fixed taxonomy adapted from~\citet{aral2026riseaisearchimplications}. The same label was assigned to the corresponding Chinese translation to preserve cross-lingual alignment. To assess annotation accuracy, one author manually reviewed \num{60} randomly sampled assignments, of which 96.7\% were judged correct. The annotation prompt, category definitions, and illustrative examples are provided in Appendix~\ref{app:topic-guidelines}.

Both the original English queries and their Simplified Chinese translations were submitted to Baidu and Google, yielding four platform-language settings: Baidu Chinese, Baidu English, Google Chinese, and Google English. Data were collected between June 20 and July 10, 2026, using Google Chrome from Milan, Italy. The platform-language settings were collected in parallel over the same period, so matched query conditions were observed within approximately the same collection window. Google searches were performed while logged in to a Google account. The Chrome interface and Google Search language were set to English, whereas Baidu was accessed through its Chinese interface.
No incognito mode, VPN, or proxy was used, and the same environment settings were kept fixed throughout the study. Search-result collection was automated using Playwright. Only source references visible in the rendered overview were recorded; no additional source panels were manually expanded. Queries were submitted sequentially, and the rendered search page, overview status, overview text, and visible source references were recorded for each query. Overall, we collected \num{1998} query-platform observations: \num{500} Baidu Chinese records, \num{500} Baidu English records, \num{499} Google Chinese records, and \num{499} Google English records. All four platform-language settings are included in the overall overview
appearance analysis. Since Baidu English yielded only  \num{20} successful AI overviews out of \num{500} queries, we exclude it from the topic-level and source-oriented analyses and report these cases descriptively in Appendix~\ref{app:baidu-en}.

Each record stores the query, topic label, collection status, and overview text when available. A record is labeled \textit{Success} when the platform-specific extraction procedure detects a qualifying AI-overview component and extracts non-empty overview text. A qualifying overview was defined as a platform-specific AI-generated overview component detected in the rendered search-results page with non-empty generated answer text. The remaining records are classified as \texttt{NoOverview}, \texttt{Blocked}, or \texttt{Error}. \texttt{NoOverview} indicates that the results page was successfully rendered but no qualifying overview was detected, whereas \texttt{Blocked} and \texttt{Error} denote access restrictions and technical failures, respectively. Definitions, outcome counts, and representative examples are provided in Appendix~\ref{app:nonoverview}. 

Figure~\ref{fig:query-topic-distribution} shows the topic
distribution of the \num{500} base queries. The distribution is
highly imbalanced: General Knowledge contains \num{325} queries, whereas
Internet/Technology/Media contains only \num{10}, and the remaining
categories contain between \num{30} and \num{51} queries. Consequently,
topic-level percentages for the smaller categories are based on
relatively few observations and may be sensitive to individual
records. We therefore report these results as descriptive
breakdowns of the collected sample rather than as balanced
comparisons across topics.

\begin{figure}[t]
    \centering
    \includegraphics[width=\columnwidth]{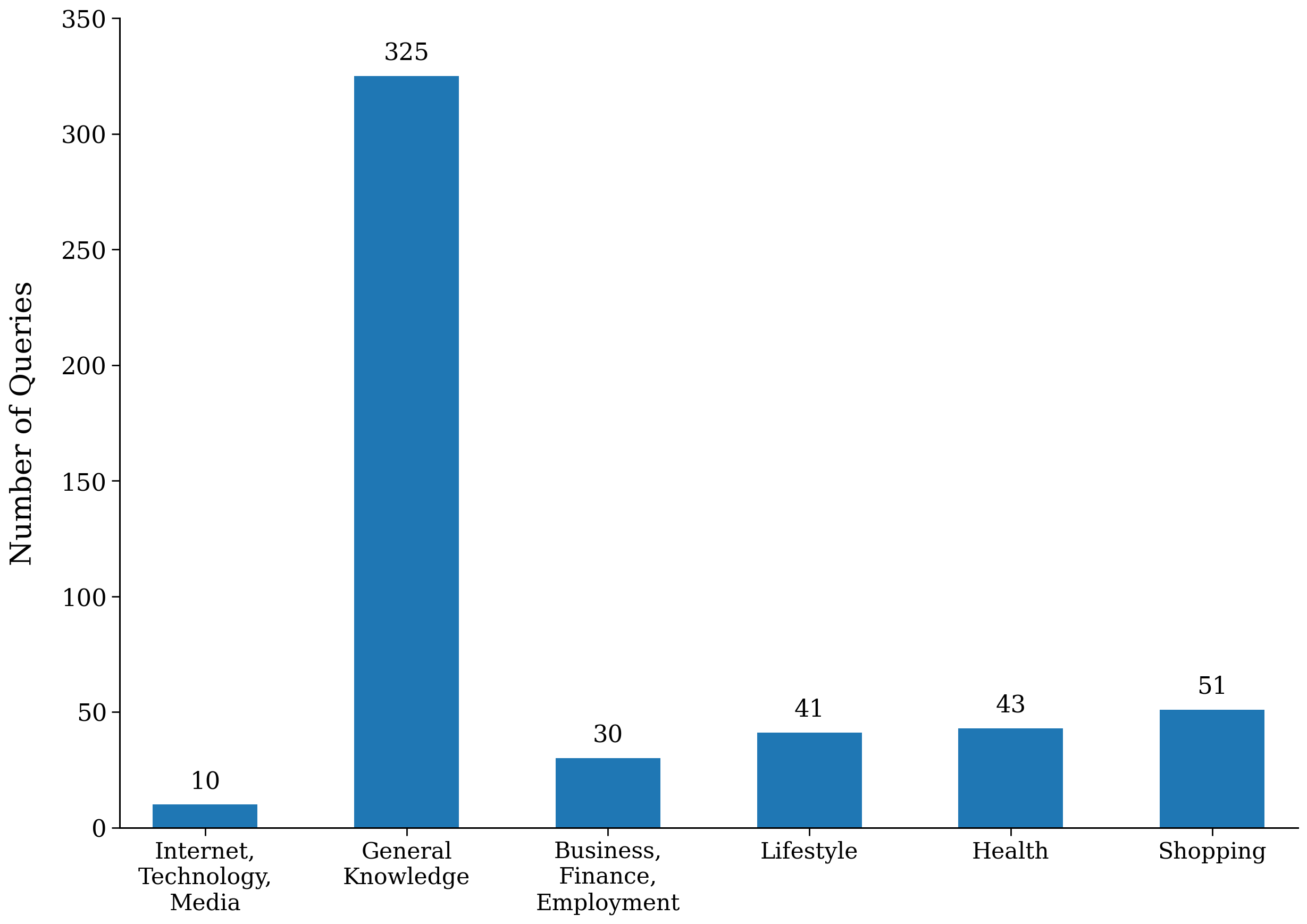}
    \caption{Topic distribution of the \num{500} base queries.}
    \label{fig:query-topic-distribution}
\end{figure}

\subsection{Source Extraction}

For successful overview records, we extract visible source references and normalize them to lowercase host domains, such as \texttt{baike.baidu.com} or \texttt{en.wikipedia.org}. We group repeated references by host while treating distinct subdomains, such as \texttt{baike.baidu.com}, \texttt{zhidao.baidu.com}, and \texttt{baijiahao.baidu.com} as separate hosts. Repeated appearances of the same host within an overview were deduplicated, so each host contributes at most one occurrence per overview. Accordingly, all reference counts reported below represent overview-level host occurrences rather than raw source-card or citation counts.
We retain these subdomains separately because
they correspond to distinct user-facing source labels and content
services in the search interface. Accordingly, the results are
interpreted at the host-domain level rather than at the level of
registrable domains or corporate ownership.

We focus on host-domain exposure  rather than exact URL exposure because the search interfaces do not always display full URLs consistently. Since Baidu and Google display source information differently, we use platform-specific extraction procedures. For Baidu, we extract displayed domains, primarily from the explicit source lists in the captured text. For Google, we map source labels and visible cards to host domains whenever possible. Therefore, cross-platform comparisons capture visible source exposure rather than providing a complete reconstruction of  the search engines' internal evidence sets. To assess extraction accuracy, one author manually reviewed \num{60} randomly sampled successful overview records, including \num{20} from each retained source-analysis setting: Baidu Chinese, Google Chinese, and Google English. Extracted host domains exactly matched the visible source references in \num{58} of the \num{60} records, corresponding to an exact-match rate of \num{96.7}\%. Detailed validation results are provided in Appendix~\ref{app:extraction-validation}.


\subsection{Metrics}

We define the overview appearance rate as
\begin{equation}
    \mathrm{OAR}
    =
    \frac{\text{number of successful overview records}}
    {\text{number of collected records}}.
\end{equation}

For each source-analysis setting, source volume is measured by the total number of overview-level host occurrences and the number of unique host domains. Let $c_d$ denote the number of overview records in which host domain $d$ appears. Its reference share is
\begin{equation}
    s_d = \frac{c_d}{\sum_i c_i},
\end{equation}
where the denominator sums the reference counts over all host domains
observed in the corresponding platform-language setting. We report the largest host-domain share (Top-1) and the Gini
coefficient of the host-domain frequency distribution, following
prior work on source concentration in AI search
systems~\citep{yang2025newssourcecitingpatterns}. Higher values
indicate more concentrated source exposure. We quantify host-domain overlap between settings using Jaccard
similarity:
\begin{equation}
    J(A,B)
    =
    \frac{|D_A \cap D_B|}{|D_A \cup D_B|},
\end{equation}
where $D_A$ and $D_B$ denote the sets of unique host domains observed
under settings $A$ and $B$~\citep{grossman2026generativeaidisruptssearch}.

As a complementary answer-level measure, we compute the cosine
similarity between overview-text embeddings. To ensure comparability across the three pairwise analyses, we retain
the common set of query intents for which all three settings generated
an overview. The overview texts are encoded using
\path{sentence-transformers/paraphrase-multilingual-MiniLM-L12-v2}
~\citep{reimers-gurevych-2020-making}. The encoder uses its default maximum sequence length of 128 tokens; longer overview texts are truncated during encoding.

For a matched query $q$ and settings $A$ and $B$, cosine similarity is
defined as
\begin{equation}
\mathrm{Sim}_{q}(A,B)
=
\frac{
\mathbf{e}_{A,q} \cdot \mathbf{e}_{B,q}
}{
\|\mathbf{e}_{A,q}\|
\|\mathbf{e}_{B,q}\|
},
\end{equation}
where $\mathbf{e}_{A,q}$ and $\mathbf{e}_{B,q}$ denote the corresponding overview-text embeddings. Higher values indicate greater semantic similarity but do not imply factual correctness, citation support, or source overlap. Because host-domain Jaccard is computed over setting-level aggregate host sets, whereas semantic similarity is computed over matched query pairs, we treat the two metrics as complementary descriptive measures. We therefore do not interpret their coexistence as evidence of a query-level relationship between answer similarity and source overlap.

\section{Results}

\subsection{Overview Appearance Rate}

Table~\ref{tab:oar} reports the overall overview appearance rates across the four platform-language settings.  The \textit{Other} category combines
\texttt{NoOverview}, \texttt{Blocked}, and \texttt{Error} records; their definitions and disaggregated counts are provided in Appendix~\ref{app:nonoverview}. 

Google generated overviews for \num{97.6}\% of Chinese queries and \num{97.0}\% of English queries. The corresponding rates for Baidu were \num{82.0}\% and \num{4.0}\%, respectively. Overview appearance therefore varied substantially across the observed platform--language settings, with the largest contrast occurring between the Baidu Chinese and Baidu English settings. As a sensitivity check, excluding Blocked and Error records from the denominator yielded corresponding OARs of \num{82.0}\%, \num{4.0}\%, \num{97.8}\%, and \num{97.6}\%, respectively, leaving the overall pattern unchanged.

\begin{table}[ht!]
\centering
\small
\begin{tabular}{lrrrr}
\toprule
Setting & Total & Success & Other & OAR \\
\midrule
Baidu Chinese & 500 & 410 & 90 & 82.0\% \\
Baidu English & 500 & 20 & 480 & 4.0\% \\
Google Chinese & 499 & 487 & 12 & 97.6\% \\
Google English & 499 & 484 & 15 & 97.0\% \\
\bottomrule
\end{tabular}
\caption{Overview appearance rate across platform-language settings.}
\label{tab:oar}
\end{table}

\begin{figure}[t]
    \centering
    \includegraphics[width=\columnwidth]{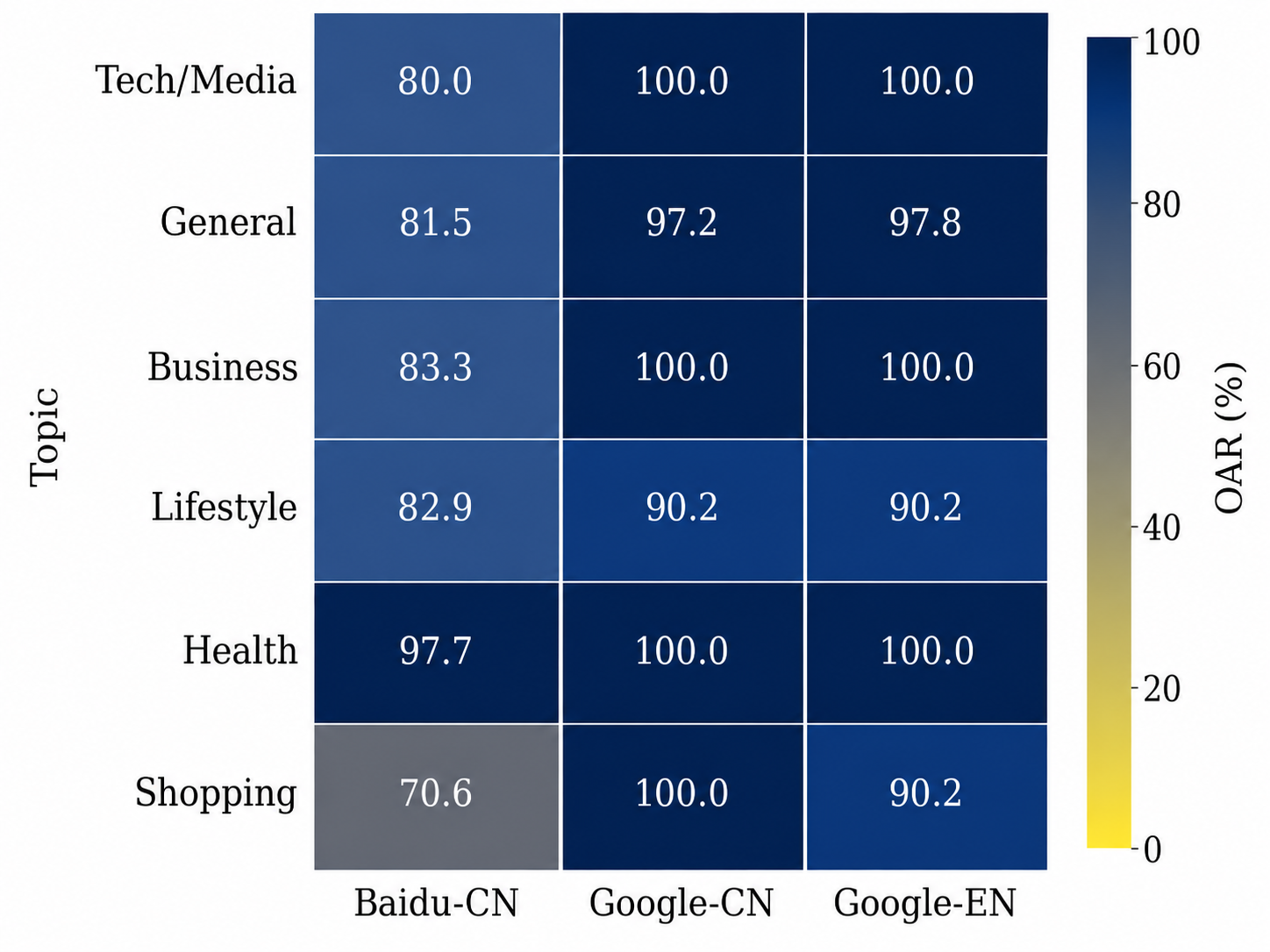}
    \caption{Topic-level overview appearance rates for Baidu Chinese, Google Chinese, and Google English. Each cell reports the percentage of queries within a topic that produced an AI overview.}
    \label{fig:oar-topic}
\end{figure}

Baidu exhibited a pronounced language difference, generating overviews for \num{82.0}\% of Chinese queries but only \num{4.0}\% of
English queries. Because Baidu English produced only \num{20} successful overviews, we retain it in the overall appearance analysis but exclude it from the topic-level and source-oriented analyses.
Figure~\ref{fig:oar-topic} therefore reports topic-level results for Baidu Chinese, Google Chinese, and Google English.

At the topic level, Google maintained high overview appearance rates across all categories, ranging from \num{90.2}\% to \num{100.0}\% in both language settings. Baidu Chinese exhibited greater variation, with the highest overview appearance rate for Health (\num{97.7}\%) and the lowest rate for Shopping (\num{70.6}\%). Rates for the remaining topics ranged from \num{80.0}\% to \num{83.3}\%.


\subsection{Source Volume and Concentration}

\begin{figure}[t]
    \centering
    \includegraphics[width=\columnwidth]{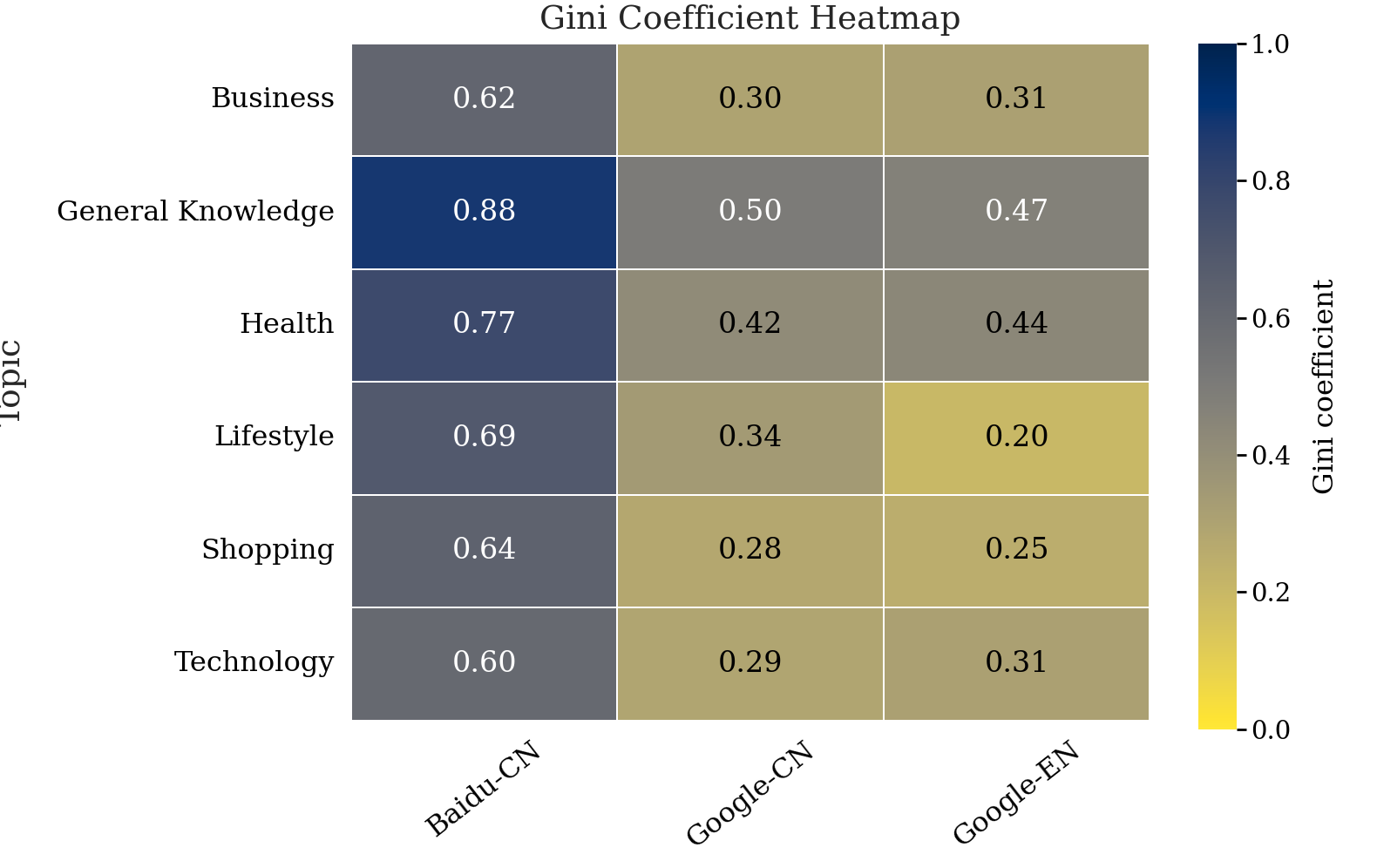}
    \caption{Topic-dependent source concentration measured by host-domain Gini coefficients. Baidu Chinese remains more concentrated than Google Chinese and Google English across most topics.}
    \label{fig:gini-topic}
\end{figure}

Detailed source-volume and host-domain concentration statistics
across the three source-analysis settings are reported in
Appendix~\ref{app:source-concentration},
Table~\ref{tab:source-conc}, while
Figure~\ref{fig:gini-topic} reports the corresponding topic-level
Gini coefficients.

Baidu Chinese contained fewer overview-level host occurrences and unique host domains than the two Google settings. After normalizing by the number of successful overview records, the mean number of distinct host occurrences per overview was \num{15.0} for Baidu Chinese, \num{18.6} for Google Chinese, and \num{16.7} for Google English. Thus, the lower aggregate host-occurrence count for Baidu Chinese was also reflected at the per-overview level, although the difference was smaller after normalization. However, its source exposure was
substantially more concentrated: its Gini coefficient was
\num{0.895}, compared with \num{0.500} for Google Chinese and
\num{0.462} for Google English. Its Top-1 host-domain share was also
higher (\num{27.1}\%, compared with \num{6.6}\% and \num{5.5}\%,
respectively).

The topic-level results show a similar pattern, with Baidu Chinese
exhibiting higher host-domain concentration than Google Chinese and
Google English across most topics.


\subsection{Top Referenced Host Domains}

\begin{figure*}[t]
    \centering
    \includegraphics[width=0.98\textwidth]{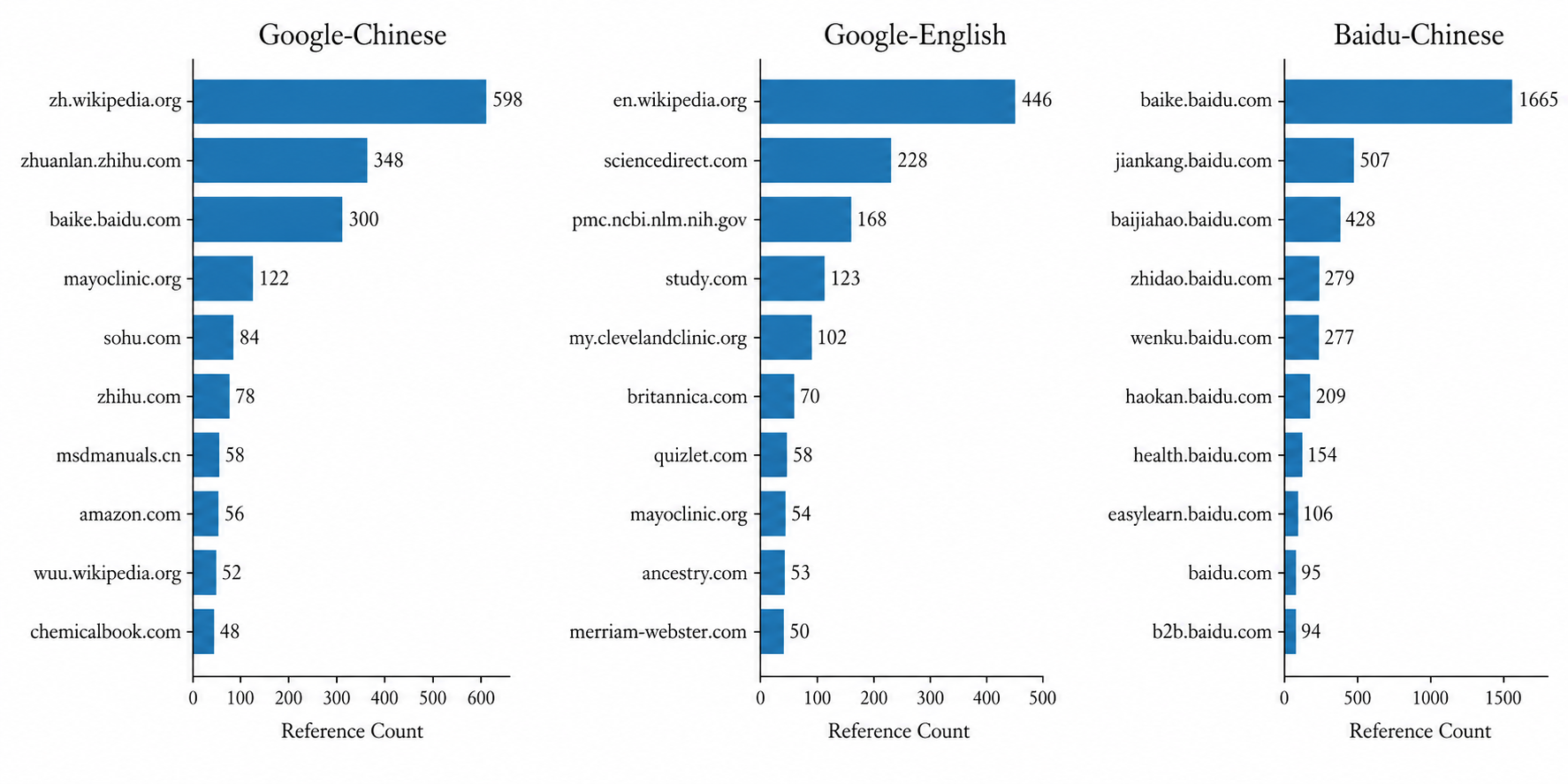}
    \caption{Overall top-referenced host domains across source-analysis settings. Baidu Chinese is dominated by Baidu-owned host domains, while Google Chinese and Google English expose a more mixed set of sources. Note that the x-axis scales differ across panels because reference counts differ substantially across settings.}
    \label{fig:top-domains}
\end{figure*}

\begin{figure}[ht!]
    \centering
    \includegraphics[width=1\linewidth]
    {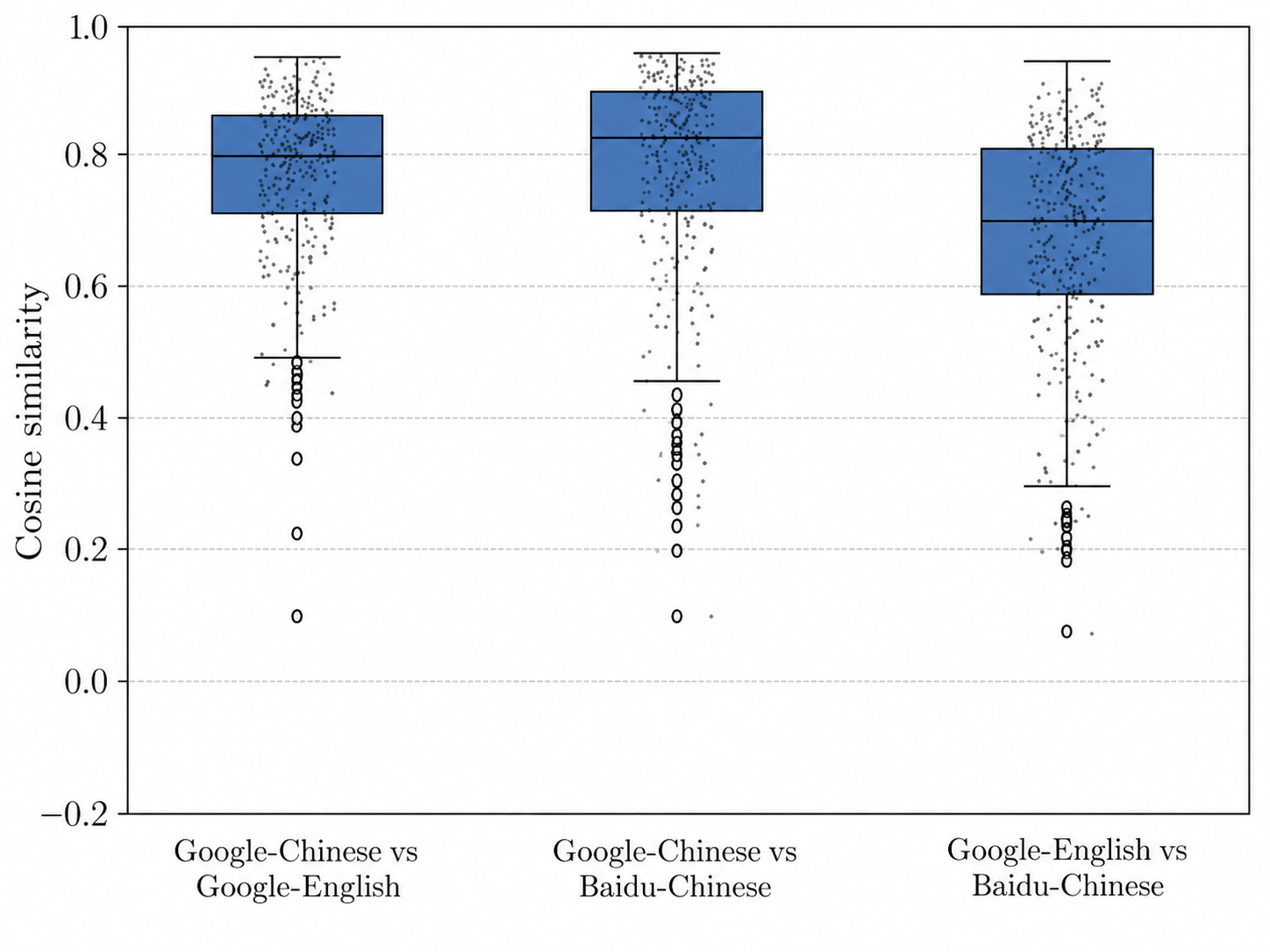}
    \caption{Query-level cosine similarity between generated answers
for matched query intents.
Only the common set of query intents for which all three settings
successfully generated an overview is included.  G-CN, G-EN, and B-CN denote Google Chinese,
    Google English, and Baidu Chinese, respectively.
    Boxes show the interquartile range, center lines
    indicate medians, and points represent matched
    queries.}
    \label{fig:semantic-similarity}
\end{figure}

Figure~\ref{fig:top-domains} shows the most frequently extracted
host domains in each source-analysis setting. For Baidu Chinese, the most frequently referenced host domains include \texttt{baike.baidu.com}, \texttt{baijiahao.baidu.com}, and \texttt{zhidao.baidu.com}. These are distinct host domains under baidu.com, while the overall host-domain frequency distribution is substantially more concentrated than in the two Google settings. 

Google Chinese and Google English exhibit a more diverse source
composition.  Their most frequently referenced domains include
\texttt{zh.wikipedia.org} and \texttt{mayoclinic.org} in Chinese,
and \texttt{en.wikipedia.org} and \texttt{sciencedirect.com} in
English. Overall, these frequency distributions are consistent with the concentration results reported above: visible source exposure is more concentrated for Baidu Chinese and more dispersed across external hosts for both Google settings.

\subsection{Cross-Platform Domain Overlap}
\label{sec:domain-overlap}

Table~\ref{tab:jaccard} reports Jaccard similarity between extracted host-domain sets. Overlap is low across all three comparisons.
Google Chinese and Google English have the highest similarity
(\num{0.034}), followed by Baidu Chinese and Google Chinese
(\num{0.032}). Baidu Chinese and Google English have the lowest
similarity (\num{0.008}). Thus, the aggregate host-domain inventories exposed by the three settings show little overlap.

\begin{table}[t]
\centering
\small
\begin{tabular}{lrrr}
\toprule
Pair & Shared & Union & Jaccard \\
\midrule
G-CN / G-EN & 179 & 5,341 & 0.034 \\
B-CN / G-CN & 103 & 3,183 & 0.032 \\
B-CN / G-EN & 25 & 3,095 & 0.008 \\
\bottomrule
\end{tabular}
\caption{Host-domain overlap between source-analysis settings. B-CN, G-CN, and G-EN denote Baidu Chinese, Google Chinese, and Google English.}
\label{tab:jaccard}
\end{table}

Jaccard similarity captures set overlap but does not account for domain frequency. For example, \texttt{baike.baidu.com} appears in both Baidu Chinese and Google Chinese, although the two settings share
only \num{103} of their \num{3183} combined host domains. Thus, individual domains may be shared even when the overall overlap between the two domain sets is limited. These Jaccard values should be interpreted as descriptive measures of aggregate host-set overlap. Because the metric is unweighted and sensitive to the number of low-frequency hosts in each setting, low values do not by themselves establish unusually low overlap relative to a statistical null model.

\subsection{Semantic Similarity of Generated Answers}
\label{sec:semantic-similarity}

Figure~\ref{fig:semantic-similarity} reports the
query-level cosine similarity between generated answers
for three pairwise comparisons: Google Chinese
versus Google English, Google Chinese versus Baidu
Chinese, and Google English versus Baidu Chinese. To ensure comparability across the three pairwise analyses, we restrict the analysis to the common set of 331 query intents for which all three settings successfully generated an overview.

The three comparisons yield median cosine similarities ranging from \num{0.701} to \num{0.813}, although their distributions differ. Google Chinese and Baidu Chinese show the highest median cosine similarity at \num{0.813} (IQR: \num{0.709}--\num{0.886}), followed by Google Chinese and Google English at \num{0.791} (IQR: \num{0.707}--\num{0.850}). Google English and Baidu Chinese show a lower median similarity of \num{0.701} (IQR: \num{0.589}--\num{0.803}) and a wider distribution, suggesting greater answer-level variation when both the platform and query language differ.

At the aggregate level, these answer-similarity results coexist with the low host-domain Jaccard similarities reported in Section~\ref{sec:domain-overlap}. Because the two metrics operate at different levels of aggregation, we do not interpret this pattern as evidence of a query-level relationship between semantic similarity and source overlap. Instead, the results show that answer similarity and aggregate source exposure capture distinct dimensions of AI-mediated search.

\section{Discussion and Conclusion}

AI-generated overviews shift search interfaces from ranked lists of
webpages toward answer-first presentations that also determine which
sources remain visible to users. Our results show that this shift is not uniform across the observed platform–language settings. Overview availability,
source concentration, and visible source overlap vary substantially
across settings, even when the queries represent the same underlying
information needs. The central takeaway is therefore methodological: answer content and visible source exposure provide complementary views of AI-mediated search and should not be treated as interchangeable indicators.

\textbf{RQ1: \textit{How frequently do AI overviews appear on Baidu
and Google under Chinese- and English-query settings?}}
Google generated overviews for nearly all collected queries, with
appearance rates of \num{97.6}\% for Chinese queries and \num{97.0}\% for English
queries. Baidu showed a substantially stronger language difference:
its appearance rate was \num{82.0}\% for Chinese queries but only \num{4.0}\% for
English queries. These results show that overview availability varied substantially across the observed platform–language settings. The very small number of
successful Baidu English cases also means that this setting cannot
support reliable topic-level or source-oriented comparisons.

\textbf{RQ2: \textit{Which host domains are displayed in
Chinese-language AI overviews on Baidu and Google, and how concentrated
is their exposure?}}
Baidu Chinese displayed fewer overview-level host occurrences and unique host domains than Google Chinese, while its source exposure was considerably more concentrated. Its host-domain-level Gini coefficient was \num{0.895}, compared with \num{0.500} for Google Chinese, and its most frequently referenced host accounted for \num{27.1}\% of all host occurrences, compared with \num{6.6}\% for Google Chinese. The most visible Baidu Chinese sources were also primarily Baidu-owned subdomains. In contrast, Google Chinese exposed a broader mixture of encyclopedic, medical,
commercial, and other external host domains. These findings show that
displaying multiple source references does not necessarily imply broad
source diversity.

\textbf{RQ3: \textit{How do visible host-domain overlap and
answer-level semantic similarity vary across Baidu Chinese, Google
Chinese, and Google English?}}

Host-domain overlap was low across all three pairwise comparisons. The
largest Jaccard similarity was \num{0.034} between Google Chinese and
Google English, whereas Baidu Chinese and Google English had a similarity
of \num{0.008}.  In contrast, matched-query answer comparisons yielded median cosine similarities ranging from \num{0.701} to \num{0.813}. Because these measures operate at different levels of aggregation, the results should not be interpreted as evidence of a query-level relationship between answer similarity and source overlap. Instead, they show that answer-level similarity and aggregate source exposure capture distinct dimensions of AI-mediated search.

More broadly, these findings show that answer-level similarity alone does not provide a complete account of AI-mediated search. Across the observed settings, median matched-query cosine similarity ranged from
\num{0.701} to \num{0.813}, while aggregate host-domain inventories showed low overlap. Audits of AI search should therefore consider not only whether an overview appears and what answer it contains, but also which sources receive visibility, how concentrated that exposure is, and how these patterns vary across platforms and languages.

This distinction is particularly important for Chinese-language search, where Baidu Chinese showed a more concentrated host-domain distribution than Google Chinese, while several of its most frequently referenced hosts were Baidu subdomains. Future work could extend this audit using independently collected native-language queries, repeated observations across locations and time periods, and evaluations of source relevance,
credibility, claim-level support, and downstream user behavior.

\clearpage
\section{Limitations}

This study is limited by the size and timing of the collected sample. Search results may change over time, and different locations or browser states may produce different pages. The extraction process also depends on page structure. If a search engine changes how it displays overview sources, the extraction rules may need to be updated. In addition, Google searches were conducted while logged in to a Google account and with the interface language set to English, so account state and interface language may have influenced the observed results. In particular, collecting Baidu from Milan may not reproduce the search experience of users accessing the service from mainland China. The observed Baidu English overview rate may reflect regional routing, interface behavior, language support, or a combination of these factors.

Another limitation comes from query translation. The Chinese query set is produced by automatically translating the original English MS MARCO queries. Although this keeps the underlying query intents approximately aligned across languages, translation may change wording, ambiguity, or cultural context. Therefore, the cross-lingual comparison should be interpreted as a comparison between original English queries and their translated Chinese counterparts, rather than between two independently collected native query sets. The topic distribution is also highly imbalanced, so topic-level results are intended as descriptive breakdowns and should not be interpreted as equally precise comparisons across categories.

The host-domain-level analysis does not judge whether a source is correct, relevant, or trustworthy. A host domain can be frequently cited while still providing weak support for a specific overview statement. Conversely, a low-frequency source may be highly relevant for a particular query. Therefore, reference frequency should be interpreted as a visibility measure rather than a quality measure. Because the analysis preserves user-facing subdomains as distinct hosts, organization-level aggregation could yield different
concentration and overlap estimates, particularly for platforms
that expose multiple services under the same corporate domain. The reported Gini and Jaccard values should therefore be interpreted as descriptive statistics under our host-domain definition. Alternative aggregation units, such as registrable domains or corporate ownership, as well as alternative overlap or concentration measures, could produce different estimates. Future work should examine the robustness of these patterns under such alternative specifications. 




\section*{Ethics Statement}

This study analyzes publicly visible search-engine interface outputs
and does not involve human participants or personally identifiable
information. The queries are drawn from the anonymized MS MARCO
dataset and are used only to compare platform behavior. All findings
are reported in aggregate.

The study does not attempt to identify individual users, infer private
system internals, or determine the cultural composition of model
training data. It does not classify websites as trustworthy or
untrustworthy, since source frequency measures interface-level
visibility rather than source quality. The findings describe the
observed collection period rather than a permanent assessment of either platform. The collection process did not attempt to bypass access restrictions or verification mechanisms; such cases were recorded as Blocked. Source-exposure measurements may also be useful for generative engine optimization. We therefore frame the results as an audit of aggregate visibility patterns rather than as actionable guidance for manipulating source rankings or platform exposure.

\bibliography{custom}

\clearpage
\appendix

\section{Source-Extraction Validation}
\label{app:extraction-validation}

We randomly sampled 60 successful overview records for manual
validation, with 20 records drawn from each retained source-analysis
setting: Baidu Chinese, Google Chinese, and Google English. For each
record, the extracted host-domain list was compared with the source
references visibly displayed in the captured overview. A record was
counted as an exact match only when all visible host domains were
correctly extracted and no unsupported domain was included.

Overall, 58 of the 60 sampled records were exact matches, yielding an
exact-match rate of 96.7\%. The remaining two records contained
extraction discrepancies. The validation was conducted by one author;
therefore, inter-annotator agreement is not applicable.

\section{Detailed Source Volume and Concentration Statistics}
\label{app:source-concentration}

Table~\ref{tab:source-conc} reports the total number of extracted
source references, the number of unique host domains, the share of
the most frequently referenced host domain, and the host-domain-level
Gini coefficient for each retained source-analysis setting.

\begin{table}[t]
\centering
\small
\setlength{\tabcolsep}{4pt}
\renewcommand{\arraystretch}{1.1}

\begin{tabular}{lrrrr}
\toprule
Setting & Refs. & Hosts & Top-1 & Gini \\
\midrule
Baidu Chinese  & 6,154 & 443   & 27.1\% & 0.895 \\
Google Chinese & 9,053 & 2,843 & 6.6\%  & 0.500 \\
Google English & 8,061 & 2,677 & 5.5\%  & 0.462 \\
\bottomrule
\end{tabular}

\caption{Source volume and concentration at the host-domain level.
Baidu English is excluded because it produced only 20 successful
overview records.}
\label{tab:source-conc}
\end{table}

\section{Classification of Non-Successful Records}
\label{app:nonoverview}

Records that were not labeled \textit{Success} were assigned to three
mutually exclusive categories: \textit{NoOverview}, \textit{Blocked},
and \textit{Error}. The distribution of these outcomes across the four
platform-language settings is reported in
Table~\ref{tab:nonoverview_breakdown}.

\textit{NoOverview} indicates that the search-results page was
successfully retrieved and rendered, but no qualifying AI-overview
component was detected. This category therefore represents an observed
absence of an overview rather than a data-collection failure.

\textit{Blocked} indicates that access to the search-results page was
restricted, preventing reliable observation of whether an AI overview
was present.

\textit{Error} indicates that the record could not be reliably
classified because of a technical failure during data collection. The
only observed error occurred when page loading exceeded the configured
25-second timeout.

Thus, \textit{NoOverview} represents an observed absence of an AI
overview, whereas \textit{Blocked} and \textit{Error} represent cases
in which overview presence could not be reliably determined.

\begin{table}[t]
\centering
\scriptsize
\setlength{\tabcolsep}{3pt}
\renewcommand{\arraystretch}{1.12}

\begin{tabular}{lrrrrr}
\toprule
Setting & Success & NoOvr. & Blocked & Error & Total \\
\midrule
Baidu Chinese  & 410 & 90  & 0 & 0 & 500 \\
Baidu English  & 20  & 479 & 0 & 1 & 500 \\
Google Chinese & 487 & 11  & 1 & 0 & 499 \\
Google English & 484 & 12  & 3 & 0 & 499 \\
\bottomrule
\end{tabular}

\caption{Distribution of successful and non-successful outcomes across
platform-language settings. \textit{NoOvr.} denotes
\textit{NoOverview}.}
\label{tab:nonoverview_breakdown}
\end{table}

Representative page-level examples illustrate the distinction among
these categories. A \textit{NoOverview} record contains a normally
rendered search-results page with standard results but no detected
AI-overview component. A \textit{Blocked} record instead displays an
access-restriction or verification page rather than usable search
results. The single \textit{Error} record corresponds to a page that
did not complete loading within the 25-second timeout.

\section{Descriptive Results for Baidu English}
\label{app:baidu-en}

As shown in Table~\ref{tab:baidu-en-topic}, Baidu English produced 20 successful AI overviews among the 500
submitted queries. Because of this small sample, the results below
are reported descriptively and are not used for cross-setting
comparisons.

\begin{table}[ht!]
\centering
\small
\begin{tabular}{lr}
\hline
Topic & Successful cases \\
\hline
Internet/Technology/Media & 2 \\
General Knowledge & 9 \\
Business/Finance/Employment & 1 \\
Lifestyle & 1 \\
Health & 1 \\
Shopping & 6 \\
\hline
Total & 20 \\
\hline
\end{tabular}
\caption{Topic distribution of successful Baidu English overview
cases.}
\label{tab:baidu-en-topic}
\end{table}

\begin{table*}[t]
\centering
\small
\begin{tabular}{p{0.20\textwidth} p{0.34\textwidth} p{0.39\textwidth}}
\hline
\textbf{Category} & \textbf{Definition} & \textbf{Example Queries} \\
\hline

\hline
Internet/Technology/Media &
Software, hardware, digital devices, online services, internet
platforms, technology, news, and media. &
\texttt{how to transfer photos and videos from android to computer} \newline
\texttt{what is a web domain} \newline
\texttt{who invented the first mechanical computer}
\\[0.8em]

\hline
General Knowledge &
General factual or informational queries that are not more
appropriately covered by another domain-specific category. &
\texttt{what are electromagnetic waves made of} \newline
\texttt{where is the amur leopard found} \newline
\texttt{what is a apothem}
\\[0.8em]

\hline
Business/Finance\newline/Employment &
Companies, markets, banking, personal finance, careers,
occupations, and employment. &
\texttt{what qualifications do a police officer need} \newline
\texttt{average salary of a teacher in texas} \newline
\texttt{when must ira distributions begin}
\\[0.8em]

\hline
Lifestyle &
Travel, food, home, leisure, hobbies, relationships, and
everyday activities. &
\texttt{closest airport to larissa greece} \newline
\texttt{how long to steam artichokes in steamer} \newline
\texttt{weather in oregon in may}
\\[0.8em]

\hline
Health &
Diseases, symptoms, diagnosis, treatment, medicine,
healthcare, and wellbeing. &
\texttt{what causes inflamed pancreas} \newline
\texttt{how often should men get a PSA test} \newline
\texttt{what types of bacteria are found in the mouth}
\\[0.8em]

\hline
Shopping &
Products, brands, prices, purchasing decisions,
availability, and product comparisons. &
\texttt{is it better to buy a car from a dealer or owner} \newline
\texttt{average cost for custom canopy} \newline
\texttt{cost to replace brakes}
\\

\hline
\end{tabular}

\caption{Topic-category definitions and example queries. All examples
are drawn from the English query dataset used for topic annotation and
are reproduced verbatim, including any grammatical errors in the
original queries.}

\label{tab:topic-definitions}
\end{table*}

\section{Topic Annotation Procedure}
\label{app:topic-guidelines}

All topic labels were generated automatically using GPT-5.5 through
the ChatGPT interface. Human involvement was limited to validating a random sample of the automatically assigned labels and correcting the two mismatched labels
identified during validation. Each original English query was assigned exactly one topic label. The
same label was then assigned to its Simplified Chinese translation,
rather than annotating the translated query independently, to preserve
cross-lingual alignment.

\subsection{Annotation Prompt}
\label{app:topic-prompt}

The same prompt and category taxonomy were used for all queries. The
model was instructed to identify the primary information need and
return exactly one of the six permitted category labels.

The following prompt template was used:

\begin{quote}
\small
\raggedright

Assign the following search query to exactly one of the categories
below according to its primary information need.

\begin{itemize}
    \item Internet/Technology/Media
    \item General Knowledge
    \item Business/Finance/Employment
    \item Lifestyle
    \item Health
    \item Shopping
\end{itemize}

Select the single best-fitting category. If more than one category
appears relevant, choose the category that most directly represents
the user's main information need. Return only the category label and
do not provide an explanation.

\textbf{Search query:} \texttt{\{query\}}
\end{quote}

In each classification request, \texttt{\{query\}} was replaced by the
actual English query being annotated.
















\subsection{Category Definitions and Examples}
\label{app:topic-definitions}

The six categories were operationalized as shown in
Table~\ref{tab:topic-definitions}. The examples are drawn from the
query dataset and illustrate the primary information need associated
with each category.

\subsection{Manual Validation}
\label{app:topic-validation}

To assess the accuracy of the automatic topic annotation, one author
manually reviewed a random sample of \num{60} labeled English queries.
For each sampled query, the reviewer assessed whether the assigned
category correctly represented its primary information need according
to the definitions above.

Among the \num{60} reviewed assignments, \num{58} were judged correct,
corresponding to an accuracy of 96.7\%. The remaining two assignments
were considered mismatched and were corrected before downstream
analysis. Because the validation was conducted by one author,
inter-annotator agreement was not computed.
\end{document}